\documentstyle[amssymb,preprint,aps]{revtex}

\begin{document}
\title{{\bf Bound Energy Masses of Mesons Containing the Fourth Generation and
Iso-singlet Quarks }}
\author{Sameer M. Ikhdair\thanks{%
Electronic address (internet): sameer@neu.edu.tr} and \ Ramazan Sever\thanks{%
Electronic address (internet): sever@metu.edu.tr}}
\address{$^{\ast }$Department of Electrical and Electronic Engineering, Near East\\
University, Nicosia, North Cyprus\\
$^{\dagger }$Department of Physics, Middle East Technical University, 06530\\
Ankara,Turkey\\
PACS NUMBER(S): 12.39.Pn, 14.40-n, 12.39.Jh, 14.65.-q, 13.30.Gd
Shifted large N-expansion method, Quarkonia, Mesons}
\date{\today
}
\maketitle

\begin{abstract}
The fourth Standard Model (SM) family quarks and weak iso-singlet quarks
predicted by ${\rm E}_{6\text{ }}$ GUT are considered. The spin-average of
the pseudoscalar $\eta _{4}(n^{1}S_{0})$ and vector $\psi _{4}(n^{3}S_{1})$
quarkonium binding masses of the new mesons formed by the fourth Standard
Model (SM) family and iso-singlet ${\rm E}_{6\text{ }}$ with their mixings
to ordinary quarks are investigated. Further, the fine and hyperfine mass
splittings of the these states are also calculated. We solved the
Schr\"{o}dinger equation with logarithmic and Martin potentials using the
Shifted large-${\rm N}$ expansion technique. Our results are compared with
other models to gauge the reliability of the predictions and point out
differences.
\end{abstract}

\begin{center}
$\bigskip $
\end{center}

\begin{verbatim}

\end{verbatim}

\section{INTRODUCTION}

\noindent The toponium quark does not form hadronic states because
of its large mass value ($m_{t}\approx 175$ $GeV)$ and full
strength of ${\rm tbW}$ vertex. On the other hand, there are
strong reasons that the fourth Standard Model (SM) family should
exist [1,2]. The flavor democracy (Democratic Mass Matrix
approach) [2] favors the existence of the nearly degenerate fourth
SM family, whereas the fifth SM family is disfavored both by the
mass phenomenology and precision tests of the ${\rm SM}$ [3]. The
fourth ${\rm SM} $ family fermions and also the isosinglet quarks
production at $\mu ^{+}\mu ^{-}$ colliders have been investigated
[3]. Thus, SM may be treated as an effective theory of fundamental
interactions rather than fundamental
particles. The multi-hundreds ${\rm GeV}$ fourth generation up-type quark $%
(u_{4}),$ if exist, will be produced at the CERN Large Hadron Collider ${\rm %
(LHC)}$ [3,4] via gluon-gluon fusion [4]. Hence, the observation of the
fourth SM family quark in ${\rm ATLAS}$ has been considered in [4,5]. It is
expected that the masses of the fourth family quarks lies between $300\leq
m_{u_{4}}\leq 700$ $GeV$ with preferable value $m_{u_{4}}=4gw\eta
=8m_{w}\cong 640$ $GeV$\ [3]. The partial-wave unitarity leads to $m_{Q}\leq
700GeV=4m_{t}$ and in general $m_{t}\ll m_{u_{4}}\ll m_{5}.$ The fourth
generation up-type quark would predominatly decay via $u_{4}\rightarrow Wb,$
with an expected event topologies are similar to those for $t-$quark pair
production. The best channel for observing it will be [5]: $u_{4}\overline{u}%
_{4}\rightarrow WWb\overline{b}\rightarrow (l\nu )(jj)(b\overline{b}),$
where one $W$ decays leptonically and the other hadronically. The mass
resolution is estimated to be $20(40)$ ${\rm GeV}$ for $m_{u_{4}}=320(640)$ $%
{\rm GeV.}$ The pseudoscalar quarkonium state $\eta _{4}(^{1}S_{0})$ formed
by the ${\rm SM}$ fourth generation quarks is the best candidate among the
fourth generation quarkonia to be produced at the ${\rm LHC}$ and ${\rm VLHC}
$ [6]. On the other hand, new heavy quarks known as weak iso-singlet quarks
are also predicted by various extensions of SM and by ${\rm E}_{6}$ GUT [7]
which is favored by superstring theory [8]. For down-type quark $(d_{4})$,
the decay mode is: $d_{4}\rightarrow Wt$ and the final state contains four $%
W $ bosons: $d_{4}\overline{d}_{4}\rightarrow tW^{-}\overline{t}%
W^{+}\rightarrow bW^{+}W^{-}\overline{b}W^{-}W^{+}.$

The small inter-family mixings [9] leads to the formation of the fourth
family quarkonia. The most promising candidate for ${\rm LHC}$ is the
pseudoscalar quarkonium state, $\eta _{4},$ which will be produced
reasonantly via gluon-gluon fusion particularly through decay channel: $\eta
_{4}\rightarrow ZH$ [10].

In spite that the masses of new quarks are larger than $m_{t},$ they can
form new hadrons because of the small inter-family mixings leads to the
formation of the fourth family quarkonia between new heavy and ordinary
quarks $(u,d,s,c,b).$ Indeed, according to the parametrization of mass
matrices given in [9] mixing between the fourth and third family quarks is
predicted to be $\left| V_{qu_{4}}\right| \approx 10^{-3}.$ Similar
situation is expected for iso-singlet down-type quarks. The condition for
forming $(Q\overline{Q})$ quarkonia states with new heavy quarks is [11]\

\begin{equation}
m_{Q}\leq (125\text{ }GeV)\left| V_{Qq}\right| ^{-2/3}.
\end{equation}
where $q=d,s,b$ for $Q=u_{4}$ and $q=u,c,t$ for $Q=d_{4}.$ Differing from $t$
quark, fourth family quarks will form quarkonia because $u_{4}$ and $d_{4}$
are almost degenerate and their decays are suppressed by small mixings
[2,9,12].

One of the important goals of the present work is to extend the shifted
large-${\rm N}$ expansion technique (SLNET) developed for the
Schr\"{o}dinger wave equation $\left[ 13,14,15\right] $ and then applied to
semirelativistic wave equations [16,17] to reproduce the spectroscopy of the
fourth SM generation up-type $(u_{4})$ quark and weak iso-singlet down-type $%
(d_{4})$ quark with their small mixings to the ordinary type $(u,d,s,c,b)$
quarks.

Here, we study the present status of the new heavy mesons formed by new
quarks in the framework of nonrelativistic potential model and give some
predictions for their bound energy masses. In section II, we present the
solution of the Schr\"{o}dinger equation using the flavor-independent
logarithmic and Martin potentials for the the self- and non-self conjugate $%
q_{i}\overline{q}_{i}$ and $q_{i}\overline{q}_{j}$ mass spectra,
respectively. A brief conclusion appear in section III. \

\section{New Heavy Mesons}

\noindent The fourth ${\rm SM}$ family and ${\rm E}_{6}$ isosinglet quarks
have formed hadron states if their mixing with ordinary known quarks is
sufficiently small. In the fourth family quarks, the parametrization given
in [9] satisfies condition (1), whereas new hadrons are not formed in the
case of paramerization given in [1]. Concerning ${\rm E}_{6}$ iso-singlet
quarks, we have no similar parametrization (one deals with $6\times 6$ mass
matrix) and one can make qualitative estimations. For example, if the
lightest isosinglet quark has the mass $m_{d_{4}}\cong 0.5$ $TeV$, new heavy
hadrons are formed for $\left| V_{qd_{4}}\right| <0.09.$ In order to
calculate the binding masses of new mesons we investigate two potentials:
(1) Logarithmic potential of the form [14,16,18]

\begin{equation}
V(r)=-0.6631+0.733\ln (r\times 1GeV),
\end{equation}
where $r$ is the interquark distance which is singular at $r=0$. (2)
Martin's potential of the form [14,16,19]

\begin{equation}
V(r)=-8.093+6.898r^{0.1},
\end{equation}
which behaves in some respects, as its power approaches to zero, as the
logarithmic potential. Hence, these potential types are stimulated by the
approximate equality of mass splittings [18]

\begin{equation}
M(\Upsilon ^{\prime }(b\overline{b}))-M(\Upsilon (b\overline{b}))\approx
M(\psi ^{\prime }(c\overline{c}))-M(\psi (c\overline{c})),
\end{equation}
which is independent upon reduced mass. These static quarkonium potentials
are monotone nondecreasing, and concave functions satisfying the condition
[14]

\begin{equation}
V^{\prime }(r)>0\text{ \ and }V^{\prime \prime }(r)\leq 0.
\end{equation}
For two particles system, we shall consider the ${\rm N-}$dimensional space
Schr\"{o}dinger equation for any spherically symmetric potential $V(r)$. If $%
\psi ({\bf r})$ denotes the Schr\"{o}dinger's wave function, a separation of
variables $\psi ({\bf r})=Y_{\ell ,m}(\theta ,\phi )u(r)/r^{(N-1)/2}$ gives
the following radial equation ($\hbar =c=1)$ [13,14,15,16,17]

\begin{equation}
\left\{ -\frac{1}{4\mu }\frac{d^{2}}{dr^{2}}+\frac{[\overline{k}-(1-a)][%
\overline{k}-(3-a)]}{16\mu r^{2}}+V(r)\right\} u(r)=E_{n,\ell }u(r),
\end{equation}
where $\mu =\left( m_{q_{i}}m_{q_{j}}\right) /(m_{q_{i}}+m_{q_{j}})$ is the
reduced mass for the two quarkonium composite particles. Here, $E_{n,l}$
denotes the Schr\"{o}dinger binding energy of meson, and $\overline{k}%
=N+2\ell -a,$ with $a$ representing a proper shift to be calculated later on
and $l$ is the angular quantum number. We follow the shifted $1/N$ or $1/%
\overline{k}$ expansion method [13,14,15] by defining
\begin{equation}
V(x(r_{0}))\;=\stackrel{\infty }{%
\mathrel{\mathop{\sum }\limits_{m=0}}%
}\left( \frac{d^{m}V(r_{0})}{dr_{0}^{m}}\right) \frac{\left( r_{0}x\right)
^{m}}{m!Q}\overline{k}^{(4-m)/2},
\end{equation}
and also the energy eigenvalue expansion [14,15]

\begin{equation}
E_{n,l}\;=\stackrel{\infty }{%
\mathrel{\mathop{\sum }\limits_{m=0}}%
}\frac{\overline{k}^{(2-m)}}{Q}E_{m},
\end{equation}
where $x=\overline{k}^{1/2}(r/r_{0}-1)$ with $r_{0}$ is an arbitrary point
where the Taylor's expansions is being performed about and $Q$ is a scale
parameter to be set equal to $\overline{k}^{2}$ at the end of our
calculations. Following the approach presented by Ref.[14], we give the
necessary expressions for calculating the binding energies:

\begin{equation}
E_{0}=V(r_{0})+\frac{Q}{16\mu r_{0}^{2}},
\end{equation}
\begin{equation}
E_{1}=\frac{Q}{r_{0}^{2}}\left[ \left( n_{r}+\frac{1}{2}\right) \omega -%
\frac{(2-a)}{8\mu }\right] ,
\end{equation}
\begin{equation}
E_{2}=\frac{Q}{r_{0}^{2}}\left[ \frac{(1-a)(3-a)}{16\mu }+\alpha ^{(1)}%
\right] ,
\end{equation}
\begin{equation}
E_{3}=\frac{Q}{r_{0}^{2}}\alpha ^{(2)},
\end{equation}
where $\alpha ^{(1)}$ and $\alpha ^{(2)}$ are two useful expressions given
by Imbo {\it et al }[13] and also the scale parameter $Q$ is defined by the
relation

\begin{equation}
Q=8\mu r_{0}^{3}V^{\prime }(r_{0}).
\end{equation}
Thus, for the $N=3$ physical space, the Schr\"{o}dinger binding energy to
the third order is [14]

\begin{equation}
E_{n,\ell }=V(r_{0})+\frac{1}{2}r_{0}V^{\prime }(r_{0})+\frac{1}{r_{0}^{2}}%
\left[ \frac{(1-a)(3-a)}{16\mu }+\alpha ^{(1)}+\frac{\alpha ^{(2)}}{%
\overline{k}}+O\left( \frac{1}{\overline{k}^{2}}\right) \right] .
\end{equation}
where the shifting parameter, $a$, is defined by

\begin{equation}
a=2-(2n_{r}+1)\left[ 3+\frac{r_{0}V^{\prime \prime }(r_{0})}{V^{\prime
}(r_{0})}\right] ^{1/2},
\end{equation}
and the root, $r_{0},$ is being determined via

\begin{equation}
1+2l+(2n_{r}+1)\left[ 3+\frac{r_{0}V^{\prime \prime }(r_{0})}{V^{\prime
}(r_{0})}\right] ^{1/2}=\left[ 8\mu r_{0}^{3}V^{\prime }(r_{0})\right]
^{1/2},
\end{equation}
where $n_{r}=n-1$ is the radial quantum number and $n$ is the principal
quantum number. Once $r_{0}$ is found via equation (16), then the
Schr\"{o}dinger binding energy of the $q_{i}\overline{q}_{j}$ system in (14)
becomes relatively simple and straightforward. Hence, the bound state mass
of the $q_{i}\overline{q}_{j}$ system is written as

\begin{equation}
M(q_{i}\overline{q}_{j})_{nl}=m_{q_{i}}+m_{q_{j}}+2E_{n,l}.
\end{equation}
The expansion parameter $1/N$ or $1/\overline{k}$ becomes smaller as $l$
becomes larger since the parameter $\overline{k}$ is proportional to $n$
which it appears in the denominator in higher-order correction.

Since systems\ that we investigate in the present work are often considered
as nonrelativistic system, then our treatment is based upon Schr\"{o}dinger
equation with a Hamiltonian
\begin{equation}
H_{o}=-\frac{\triangledown ^{2}}{2\mu }+V(r)+V_{SD},
\end{equation}
where $V_{SD}$ is the spin-dependent term taking the simple form (cf.
Ref.[14] and the references therein)
\begin{equation}
\text{ \ }V_{SD}\longrightarrow V_{SS}=\frac{32\pi \alpha _{s}}{%
9m_{q_{1}}m_{q_{2}}}\delta ^{3}({\bf r}){\bf s}_{1}.{\bf s}_{2}.
\end{equation}
The spin dependent potential is simply a spin-spin part and this would
enable us to make some preliminary calculations of the energies of the
lowest two S-states of the new quarks and their small mixings with the
ordinary quarks. The potential parameters in this section are all strictly
flavor-independents and are fitted to the low-lying energy levels of $c%
\overline{c}$ and $b\overline{b}$ systems. The strong coupling constant $%
\alpha _{s}$ is fitted to the observed charmonium hyperfine splitting of 117
MeV. The numerical values of $\alpha _{s}$ for the two potential types have
been adjusted in Ref.[14] as

\begin{equation}
\alpha _{s}^{(L)}(c\overline{c})\simeq 0.220\text{ \ \ and \ \ \ }\alpha
_{s}^{(M)}(c\overline{c})\simeq 0.251,
\end{equation}
which are found to be dependent on the potential type. Baldicchi and
Prosperi [20] used the standard running QCD coupling expression

\begin{equation}
\alpha _{s}({\bf Q})=\frac{4\pi }{\left( 11-\frac{2}{3}n_{f}\right) \ln
\left( \frac{{\bf Q}^{2}}{\Lambda ^{2}}\right) }.
\end{equation}
with $n_{f}=4$ and $\Lambda =0.2~GeV$ cut at a maximum value $\alpha
_{s}(0)=0.35,$ to give the right $J/\psi -\eta _{c}$ splitting of $c%
\overline{c}$ quarkonium and to treat properly the infrared region. Detail
on their numerical works are given in Ref. [20]. Whereas Brambilla and Vairo
[21] took in their perturbative analysis $0.26\leq \alpha _{s}(\mu
=2GeV)\leq 0.30.$

The potential parameters in (2) and (3) together with the quark masses are
obtained from the experimental SAD mass: $\overline{M}(\psi (1S))$ (cf. e.g.
Refs.[14,16]). In our calculations we used $m_{u}=m_{d}=0.367$ $GeV,$ $%
m_{s}=0.561$ $GeV$ whereas $m_{c}$ and $m_{b}$ masses are given by Ref.[14].
Further, the fourth SM family up-type quark and $E_{6}$ isosinglet down-type
quark masses are taken as $m_{u_{4}}=638.6$ $GeV$ [9,22] and $m_{d_{4}}=0.5$
$TeV$ [22], respectively.

Firstly, we calculate the masses of bound states of $c\overline{c}$ and $b%
\overline{b}$ in Table 1 for the two potential types. It is seen that our
results are in good agreement with the experimental results [23,24]. In
Table 2, we present the masses of bound states of quarkonia and mesons
formed by the fourth SM family up-type quark and $E_{6}$ isosinglet quark,
respectively. In Table 3, the predicted mass splittings is compared with
Ref.[22]. In Table 4, the hyperfine mass splittings $\psi _{4}-\eta _{4}$
for the first two $S-$states are also given. The best mechanism for the
production of heavy mesons formed by $u_{4}$ and $d_{4}$ quarks is the
resonance formation of $3S$ and $4S$ quarkonia at lepton colliders with
subsequent decay into corresponding meson-antimeson states.

\section{CONCLUSION}

We have produced the mesons formed by new heavy quarks. The mass splitting
of the logarithmic potentials is found to be independent on the reduced mass
and is constant in magnitude for any chosen state for all studied quarkonium
families. However, it is seen in Martin's potential that the mass splittings
are equal for any chosen state in the fourth family, $u_{4},$ quark and
isosinglet quark with the same type of ordinary quark mixing. Further, the
hyperfine splitting mass is found to be much smaller compared to the large
mass of the new mesons. Therefore, the pseudoscalar $\eta _{4}(n^{1}S_{0})$
and the vector $\psi _{4}(n^{3}S_{1})$ have nearly same masses. On the other
hand, the Cornell potential fails to produce the masses of the $u_{4}%
\overline{u}_{4}$ and $d_{4}\overline{d}_{4}$ quarkonium famileis and their
hyperfine mass splittings properly because of the small values of the roots $%
r_{0}$ for the lowest states where the Coulombic part of the potential goes
to high order fas we take higher order derivatives in the ${\rm SLNET}$
method.

\acknowledgments
S. M. Ikhdair thanks his wife, Oyoun, and his son, Musbah, for their
assistance, patience and love.

\newpage

\baselineskip= 2\baselineskip

\bigskip \bigskip
\begin{table}[tbp]
\caption{The pseudoscalar, vector and the spin-averaged masses together with
hyperfine splittings for the first two-states of $c\overline{c}$ and $b%
\overline{b}$ states (in $MeV$).}
\label{Table 1}$
\begin{array}{ccccccc}
\tableline\tableline\text{States} & c\overline{c} & b\overline{b} & c%
\overline{c} & b\overline{b} & c\overline{c} & b\overline{b} \\
\tableline & \text{Logarithmic:} &  & \text{Martin:} &  &
\text{Experiment
[23,24]:} &  \\
1S & 3068 & 9444 & 3068 & 9445 & 3068\pm 2 & 9448\pm 5 \\
1^{3}S_{1} & 3097 & 9460 & 3097 & 9461 & 3097 & 9460 \\
1^{1}S_{0} & 2980 & 9395 & 2980 & 9397 & 2980 & - \\
\Delta _{1S} & 117 & 65 & 117 & 64 & 117 & 10017\pm 5 \\
2S & 3654 & 10030 & 3670 & 10018 & 3663\pm 5 & 10023 \\
2^{3}S_{1} & 3668 & 10037 & 3685 & 10030 & 3686 &  \\
2^{1}S_{0} & 3614 & 10008 & 3625 & 9994 & 3622\pm 12\tablenote{Here we cite
Ref.[25].} &  \\
\Delta _{2S} & 54 & 30 & 60 & 33 & 57\pm 8\tablenote{Here we cite Ref.[26].}
&  \\
3S & 3976 & 10352 & 4021 & 10351 &  & 10350\pm 5 \\
4S & 4200 & 10575 & 4272 & 10590 &  & 10580 \\
5S & 4370 & 10746 & 4469 & 10777 &  &  \\
1P & 3505 & 9881 & 3505 & 9861 & 3525\pm 1 & 9900\pm 1 \\
2P & 3878 & 10254 & 3907 & 10243 &  & 10261\pm 1
\end{array}
$%
\end{table}

\bigskip

\begin{table}[tbp]
\caption{Predicted Spin-averaged masses of the bound states formed by the
fourth SM family $u_{4}$ and isosinglet $d_{4}$ quarks \ (in $GeV$).}
\label{Table 2}
\begin{tabular}{lllllllllll}
SAD Mass & $u_{4}\overline{u}_{4}$ & $u_{4}\overline{u}$ & $u_{4}\overline{s}
$ & $u_{4}\overline{c}$ & $u_{4}\overline{b}$ & $d_{4}\overline{d}_{4}$ & $%
d_{4}\overline{u}$ & $d_{4}\overline{s}$ & $d_{4}\overline{c}$ & $d_{4}%
\overline{b}$ \\
\tableline Logarithmic: &  &  &  &  &  &  &  &  &  &  \\
$\overline{M}(1S)$ & 1275.05\tablenote{Fourth family system scale offset by
$2.05$ $GeV$ to agree with Ref.[22].} & 639.30 & 639.34 & 639.92 & 642.89 &
997.94\tablenote{Iso-singlet system scale offset by $1.96$ $GeV$ to agree
with Ref.[22].} & 500.70 & 500.74 & 501.32 & 504.29 \\
$\overline{M}(2S)$ & 1275.64 & 639.88 & 639.92 & 640.50 & 643.47 & 998.53 &
501.28 & 501.32 & 501.90 & 504.88 \\
$\overline{M}(3S)$ & 1275.96 & 640.21 & 640.24 & 640.82 & 643.80 & 998.85 &
501.61 & 501.64 & 502.22 & 505.20 \\
$\overline{M}(4S)$ & 1276.18 & 640.43 & 640.47 & 641.05 & 644.02 & 999.07 &
501.83 & 501.87 & 502.45 & 505.42 \\
$\overline{M}(1P)$ & 1275.49 & 639.73 & 639.77 & 640.35 & 643.33 & 998.38 &
501.13 & 501.17 & 501.75 & 504.73 \\
$\overline{M}(2P)$ & 1275.86 & 640.11 & 640.15 & 640.73 & 643.70 & 998.75 &
501.51 & 501.55 & 502.13 & 505.10 \\
$\overline{M}(3P)$ & 1276.11 & 640.36 & 640.39 & 640.97 & 643.95 & 999.00 &
501.76 & 501.79 & 502.37 & 505.35 \\
&  &  &  &  &  &  &  &  &  &  \\
Martin: &  &  &  &  &  &  &  &  &  &  \\
$\overline{M}(1S)$ & 1274.82 & 638.77 & 638.80 & 639.62 & 642.64 & 997.69 &
500.17 & 500.20 & 501.02 & 504.04 \\
$\overline{M}(2S)$ & 1275.28 & 639.39 & 639.42 & 640.21 & 643.20 & 998.15 &
500.79 & 500.82 & 501.61 & 504.60 \\
$\overline{M}(3S)$ & 1275.54 & 639.76 & 639.78 & 640.55 & 643.52 & 998.42 &
501.16 & 501.18 & 501.95 & 504.92 \\
$\overline{M}(4S)$ & 1275.73 & 640.02 & 640.03 & 640.79 & 643.75 & 998.61 &
501.42 & 501.43 & 502.19 & 505.15 \\
$\overline{M}(1P)$ & 1275.15 & 639.22 & 639.25 & 640.05 & 643.04 & 998.03 &
500.62 & 500.65 & 501.45 & 504.44 \\
$\overline{M}(2P)$ & 1275.46 & 639.64 & 639.66 & 640.44 & 643.41 & 998.33 &
501.04 & 501.06 & 501.84 & 504.81 \\
$\overline{M}(3P)$ & 1275.67 & 639.93 & 639.94 & 640.70 & 643.67 & 998.55 &
501.33 & 501.34 & 502.11 & 505.07
\end{tabular}
\end{table}

\bigskip \bigskip
\begin{table}[tbp]
\caption{Predicted fine-splittings of S and P levels (in $MeV$).}$
\begin{array}{ccccccccccccccccc}
\tableline\tableline\text{Mass splittings} & u_{4}\overline{u}_{4} & [22] &
u_{4}\overline{u} & [22] & u_{4}\overline{s} & [22] & u_{4}\overline{c} &
u_{4}\overline{b} & d_{4}\overline{d}_{4} & [22] & d_{4}\overline{u} & [22]
& d_{4}\overline{s} & [22] & d_{4}\overline{c} & d_{4}\overline{b} \\
\tableline\text{Logarithmic:} &  &  &  &  &  &  &  &  &  &  &  &  &  &  &  &
\\
2S-1S & 59 & 59 & 58 & 59 & 58 & 59 & 58 & 58 & 59 & 59 & 58 & 59 & 58 & 59
& 58 & 59 \\
3S-2S & 32 & 32 & 33 & 33 & 32 & 32 & 32 & 33 & 32 & 32 & 33 & 33 & 32 & 32
& 32 & 32 \\
4S-3S & 22 & 22 & 22 & 22 & 23 & 23 & 23 & 22 & 22 & 22 & 22 & 22 & 23 & 23
& 23 & 22 \\
2P-1P & 37 & - & 38 & - & 38 & - & 38 & 37 & 37 & - & 38 & - & 38 & - & 38 &
37 \\
3P-2P & 25 & - & 25 & - & 24 & - & 24 & 25 & 25 & - & 25 & - & 24 & - & 24 &
25 \\
&  &  &  &  &  &  &  &  &  &  &  &  &  &  &  &  \\
\text{Martin:} &  &  &  &  &  &  &  &  &  &  &  &  &  &  &  &  \\
2S-1S & 46 &  & 62 &  & 62 &  & 59 & 56 & 46 &  & 62 &  & 62 &  & 59 & 56 \\
3S-2S & 26 &  & 37 &  & 36 &  & 34 & 32 & 27 &  & 37 &  & 36 &  & 34 & 32 \\
4S-3S & 19 &  & 26 &  & 25 &  & 24 & 23 & 20 &  & 26 &  & 25 &  & 24 & 23 \\
2P-1P & 31 &  & 42 &  & 41 &  & 39 & 37 & 30 &  & 42 &  & 41 &  & 39 & 37 \\
3P-2P & 21 &  & 29 &  & 28 &  & 26 & 26 & 22 &  & 29 &  & 28 &  & 27 & 26
\end{array}
$%
\label{Table 3}
\end{table}

\bigskip

\mediumtext

\bigskip
\begin{table}[tbp]
\caption{Hyperfine mass splittings (in $MeV$).}$
\begin{array}{ccccccccccccc}
\tableline\tableline\text{Level} & u_{4}\overline{u}_{4} & u_{4}\overline{u}
& u_{4}\overline{s} & u_{4}\overline{c} & u_{4}\overline{b} & d_{4}\overline{%
d}_{4} & d_{4}\overline{u} & d_{4}\overline{s} & d_{4}\overline{c} & d_{4}%
\overline{b} & c\overline{c} & b\overline{b} \\
\tableline\text{Logarithmic:} &  &  &  &  &  &  &  &  &  &  &  &  \\
\Delta _{1S} & 5.69 & 0.39 & 0.48 & 0.78 & 1.39 & 6.43 & 0.49 & 0.61 & 0.99
& 1.78 & 117(117)\tablenote{The quantity in bracket is the experimental or
taken from other references.} & 65(86) \\
\Delta _{2S} & 2.62 & 0.18 & 0.22 & 0.36 & 0.64 & 2.96 & 0.23 & 0.28 & 0.46
& 0.82 & 54(57\pm 8)\tablenote{Here we cite Ref.[26].} & 30(35) \\
\text{Martin:} &  &  &  &  &  &  &  &  &  &  &  &  \\
\Delta _{1S} & 4.09 & 0.45 & 0.54 & 0.89 & 1.38 & 4.70 & 0.57 & 0.69 & 1.13
& 1.76 & 117 & 64 \\
\Delta _{2S} & 2.09 & 0.23 & 0.28 & 0.45 & 0.71 & 2.41 & 0.29 & 0.35 & 0.58
& 0.90 & 60 & 33
\end{array}
$%
\label{table1 4}
\end{table}

\bigskip \bigskip

\end{document}